\documentclass{article}
\usepackage{amsmath,amssymb}
\usepackage{mathrsfs}
\usepackage[dvips]{graphicx}
\newenvironment{ol}[1]{
\begin{enumerate}
\setlength{\partopsep}{#1pt}
\setlength{\topsep}{#1pt}
\setlength{\itemsep}{#1pt}
\setlength{\parsep}{#1pt}
\setlength{\parskip}{#1pt}}{
\end{enumerate}
}

\title{Learning to imitate stochastic time series in a compositional way by chaos}
\author{
Jun Namikawa \\
{\small Brain Science Institute, RIKEN}\\
{\small 2-1 Hirosawa, Wako-shi, Saitama, 351-0198 Japan}\\
{\small E-mail \texttt{jnamika@bdc.brain.riken.go.jp}}\\
\and
Jun Tani \\
{\small Brain Science Institute, RIKEN}\\
{\small 2-1 Hirosawa, Wako-shi, Saitama, 351-0198 Japan}\\
{\small E-mail \texttt{tani@brain.riken.go.jp}}\\
}
\date{}

\begin{document}
\maketitle
\begin{flushleft}
\end{flushleft}

\begin{abstract}
This study shows that a mixture of RNN experts model can acquire the ability to generate sequences combining multiple primitive patterns by means of self-organizing chaos.
By training of the model, each expert learns a primitive sequence pattern, and a gating network learns to imitate stochastic switching of the multiple primitives via a chaotic dynamics, utilizing a sensitive dependence on initial conditions.
As a demonstration, we present a numerical simulation in which the model learns Markov chain switching among some Lissajous curves by a chaotic dynamics.
Our analysis shows that by using a sufficient amount of training data, balanced with the network memory capacity, it is possible to satisfy the conditions for embedding the target stochastic sequences into a chaotic dynamical system.
It is also shown that reconstruction of a stochastic time series by a chaotic model can be stabilized by adding a negligible amount of noise to the dynamics of the model.
\end{abstract}

\section{Introduction}

Recurrent neural networks (RNNs) \cite{Elman90,Jordan86,Pollack91,Zipser89} are abstract models of nervous systems that can perform temporal sequence processing.
RNNs have been applied to temporal sequence learning such as sensory-motor sequence patterns \cite{Jordan86,Jordan88a}, symbolic sequences with grammar \cite{Elman90,Pollack91} and continuous spatio-temporal patterns \cite{Zipser89}.
However, in spite of the considerable amount of RNN research carried out since the mid 1980s, it has been thought that RNNs could not be scaled so as to be capable of learning complex sequence patterns, especially when the sequence patterns to be learned contain long-term dependencies \cite{Bengio94}.
This is due to the fact that the error signal cannot be propagated effectively in long-time windows of sequences using the back-propagation through time (BPTT) algorithm \cite{Rumelhart86}, because of the potential nonlinearity of the RNN dynamics \cite{Bengio94}.
How to form neural network models that can learn complex temporal sequence patterns has proved to be a challenging problem.

There have been some breakthroughs in approaches to this problem.
Echo state networks \cite{Jaeger2001,Jaeger2004} and a very similar approach, liquid state machines \cite{Maass2002}, have recently attracted considerable attention.
An echo state network possesses a large pool of hidden neurons with fixed random weights, and the pool is capable of rich dynamics.
Jaeger \cite{Jaeger2004} demonstrated that an echo state network can successfully learn the Mackey-Glass chaotic time series which is a well-known benchmark system for time series prediction.
Long Short-Term Memory (LSTM) proposed by Hochreiter and Schmidhuber \cite{Hochreiter97,Schmidhuber2002} has been investigated as a neural network architecture which is more efficient than standard RNN in terms of memory capability.
LSTM learns to bridge minimal time lags in long time steps by enforcing constant error flow through ``constant error carousels'' within special units. 
These units learn to open and close access to the constant error flow.

Tani and Nolfi \cite{Tani99} investigated the same problems, but from a different angle, focusing on the idea of compositionality for sensory-motor learning.
The term compositionality was adopted from the ``Principle of Compositionality'' \cite{Evans81} in linguistics, which claims that the meaning of a complex expression is determined by the meanings of its constituent expressions and rules used to combine them.
This principle, when translated to the context of sensory-motor learning, leads to the assertion that varied and complex patterns can be learned by adaptively combining reusable behavior primitives.
Here, acquiring behavior primitives requires a mechanism for autonomously segmenting a continuously experienced sensory-motor flow into reusable chunks.
Tani and Nolfi proposed a scheme for hierarchical segmentation of the sensory-motor flow, applying the idea of a mixture of experts \cite{Jacobs91,Jordan94,Wolpert98} to hierarchically organized RNNs.
Consider a network consisting of multiple local RNNs, organized in two levels, referred to as the base level and a higher level.
At the base level, RNNs called experts competitively learn prediction or generation of specific sensory-motor profiles.
As the winner among the experts changes, corresponding to structural changes in the sensory-motor flow, the sensory-motor flow is segmented by switching between winners.
Meanwhile, an RNN at the higher level, usually called a gating network, learns the sequence patterns of winner switching with much slower time constants than those at the base level.
The higher level learns not for details of sensory-motor patterns but for abstraction of primitive sequences with long-term dependencies.

However, how to practically perform learning at the higher level is a difficult problem, because the gating network is usually a deterministic system, whereas arbitrary composition of primitives with a certain probability distribution brings nondeterminacy to the sequence patterns of winner switching.
A possibly easy solution to this problem is to apply the output values of the gating network as a probability of winner switching, or more directly, a stochastic model such as a hidden Markov model (HMM) is used instead of the gating network.
Of course, if we use a stochastic model at the higher level, nondeterministic sequences can be learned by the model.
Nevertheless, using a stochastic model might remove from the dynamics the stability brought about by the effect of basins of attraction of RNN experts, because stochastic models such as HMMs can operate only with discrete value sequences, so a trajectory jumps discontinuously whenever a winner changes.
In particular, when sequences to be learned are continuous, for example, a sensory-motor flow, the stability might be lost by using a stochastic model, because sequence patterns of winner switching are also continuous.
An alternative solution, which is the main focus of the current paper, is to imitate nondeterminacy by using the gating network itself.
It has been reported that a nondeterministic process can be imitated by a chaotic dynamical system from the view of symbolic dynamics \cite{Hao89,Lind95}.
Moreover, Tani et al. \cite{NishimotoNN04,Tani95b} showed that an RNN can acquire the ability to imitate a nondeterministic symbolic process by experimental simulations.
The approach imitating a nondeterministic process using the gating network offers the advantage not only of preserving the dynamical stability, but also that of a consistent approach in terms of dynamical systems theory to both the base and higher levels.

In the current study, we investigate learning for a mixture of experts model focusing on the higher level, and show that the model can learn nondeterministic sequence patterns of winner switching by a self-organizing internal chaos.
In the previous study \cite{Namikawa2008}, we demonstrated that a mixture of experts model with our proposed learning method was able to learn to segment time series consisting of many primitive patterns, whereas in earlier studies such a series could not be segmented owing to near-miss problems in matching the current sequence pattern to the best expert among others which had acquired similar pattern profiles.
The previous study, however, focused on the learning process of segmenting sequences into reusable primitives with respect to spatio-temporal patterns, and so did not deal with higher level learning.
This study is therefore focused on higher level learning.

\section{Model}

\begin{figure}
\begin{center}
\scalebox{0.35}{\includegraphics{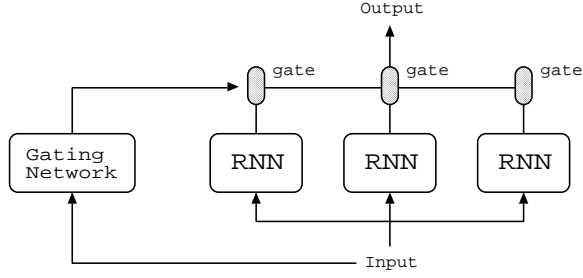}}

\caption{A system comprised of a mixture of RNN experts.}
\label{figure:mixture_of_rnn_experts}
\end{center}
\end{figure}

A mixture of RNN experts is simply a mixture of experts model for which experts are RNNs.
The mixture of RNN experts consists of expert networks together with a gating network (see Figure \ref{figure:mixture_of_rnn_experts}).
All experts receive the same input and have the same number of output neurons.
The gating network receives the past gate opening values and the input and controls gate opening.
The role of each expert is to compute a specific input-output function, and the role of the gating network is to decide which single expert is the winner on each occasion.

The dynamic states of the mixture of RNN experts at time $n$ are updated according to
\begin{equation} \label{equation:mixture_of_rnn_experts1}
\boldsymbol{u}_{n}^{(i)} = \big(1-\epsilon\big)\boldsymbol{u}_{n-1}^{(i)} + \epsilon\big( W_1^{(i)} \boldsymbol{x}_n + W_2^{(i)} \boldsymbol{c}_{n-1}^{(i)}\big),
\end{equation}
\begin{equation} \label{equation:mixture_of_rnn_experts2}
\boldsymbol{c}_n^{(i)} = \tanh(\boldsymbol{u}_{n}^{(i)} + \boldsymbol{v}_1^{(i)}),
\end{equation}
\begin{equation} \label{equation:mixture_of_rnn_experts3}
\boldsymbol{y}_n^{(i)} = \tanh(W_3^{(i)} \boldsymbol{c}_{n}^{(i)} + \boldsymbol{v}_2^{(i)}),
\end{equation}
\begin{equation} \label{equation:mixture_of_rnn_experts4}
\boldsymbol{y}_n = \sum_{i=1}^{N}g_n^{(i)} \boldsymbol{y}_n^{(i)},
\end{equation}
where $N$ is the number of experts, $\boldsymbol{x}_n$ and $\boldsymbol{y}_n$ are an input and output of the model respectively.
For each $i$, $g_n^{(i)}$, $\boldsymbol{c}_{n}^{(i)}$ and $\boldsymbol{y}_n^{(i)}$ denote the gate opening value, context and output of the expert network $i$ respectively.
Here we assume $g_n^{(i)} \geq 0$ and $\sum_{i=1}^{N}g_n^{(i)} = 1$.
The gate opening vector $\boldsymbol{g}_n$ represents the winner-take-all competition among experts to determine the output $\boldsymbol{y}_n$.
The gate opening vector $\boldsymbol{g}_n$ is the output of a gating network defined by
\begin{equation} \label{equation:mixture_of_rnn_experts5}
\boldsymbol{u}_{n}^{g} = \big(1-\epsilon^{g}\big)\boldsymbol{u}_{n-1}^{g} + \epsilon^{g}\big( W_1^{g} \boldsymbol{g}_{n-\tau} + W_2^{g} \boldsymbol{x}_{n-\tau} + W_3^{g} \boldsymbol{c}_{n-1}^{g}\big),
\end{equation}
\begin{equation} \label{equation:mixture_of_rnn_experts6}
\boldsymbol{c}_n^{g} = \tanh(\boldsymbol{u}_{n}^{g} + \boldsymbol{v}_1^{g}),
\end{equation}
\begin{equation} \label{equation:mixture_of_rnn_experts7}
\boldsymbol{b}_n = W_4^{g} \boldsymbol{c}_{n}^{g} + \boldsymbol{v}_2^{g},
\end{equation}
\begin{equation} \label{equation:mixture_of_rnn_experts8}
g_{n}^{(i)} = \frac{\exp(b_n^{(i)})}{\sum_{k=1}^N \exp(b_n^{(k)})},
\end{equation}
where $\tau$ represents a feedback time delay.
In order to satisfy $g_n^{(i)} \geq 0$ and $\sum_{i=1}^{N}g_n^{(i)} = 1$, equation (\ref{equation:mixture_of_rnn_experts8}) is given by the soft-max function.
Using the sigmoid function denoted by ${sigmoid}(x) = \frac{1}{1+\exp(-x)}$, the equation (\ref{equation:mixture_of_rnn_experts8}) can be expressed as
\begin{equation}
g_{n}^{(i)} = {sigmoid}\big(b_n^{(i)} - \ln \sum_{k \neq i} \exp(b_n^{(k)})\big).
\end{equation}
This equation indicates that the output of the gating network is given by the sigmoid function with global suppression.

\subsection{Learning Method} \label{subsection:learning_rule}

\subsubsection{Basic procedure} \label{subsubsection:basic_procedure}

We present the procedure for training the mixture of RNN experts, organized into two phases as follows:
\begin{ol}{2}
\item We compute the experts learning together with an estimate $\hat{\boldsymbol{g}}_n$ of the gate opening vector.
\item Using the estimate $\hat{\boldsymbol{g}}_n$ of gate opening vector as a target, we compute learning for the gating network.
\end{ol}
Note that in phase (1) we use the method proposed in \cite{Namikawa2008}.
This learning is a maximum likelihood estimation using the gradient descent method.
The training procedure progresses with phase (2) after convergence of phase (1).

\subsubsection{Probability distribution} \label{subsubsection:probability_distribution}

In order to define the learning method, we assign a probability distribution to the mixture of RNN experts.
Here, learnable parameters of an expert network $i$ and a gating network are denoted by $\vartheta_i = \big(W_1^{(i)}, W_2^{(i)},W_3^{(i)}, \boldsymbol{v}_1^{(i)}, \boldsymbol{v}_2^{(i)}, \boldsymbol{u}_0^{(i)}\big)$ and $\boldsymbol{\theta}^{g} = \big(W_1^{g}, W_2^{g}, W_3^{g}, W_4^{g}, \boldsymbol{v}_1^{g}, \boldsymbol{v}_2^{g}, \boldsymbol{u}_0^g\big)$, respectively.
An estimate $\hat{\boldsymbol{g}}_n$ for the gate opening vector is described by a variable $\boldsymbol{\beta}_n$ such as  
\begin{equation}
\hat{g}_n^{(i)} = \frac{\exp(\beta_n^{(i)})}{\sum_{k=1}^N \exp(\beta_n^{(k)})}.
\end{equation}

Let $X = (\boldsymbol{x}_n)_{n=1}^T$ be an input sequence, $\boldsymbol{\beta}_n$ and $\boldsymbol{\gamma}$ be parameters, where $\boldsymbol{\gamma}$ is a set of parameters given by $\gamma_i = (\vartheta_i, \sigma_i)$.
Given $X$, $\boldsymbol{\beta}_n$, and $\boldsymbol{\gamma}$, a probability density function (p.d.f.) for the output $\boldsymbol{y}_n$ is defined by
\begin{equation}
p(\boldsymbol{y}_n ~|~ X, \boldsymbol{\beta}_n, \boldsymbol{\gamma}) = \sum_{i=1}^{N} \hat{g}_n^{(i)} p(\boldsymbol{y}_n ~|~ X, \gamma_i),
\end{equation}
where $p(\boldsymbol{y}_n | X, \gamma_i)$ is given by
\begin{equation} \label{equation:mixture_of_normal_distribution}
p(\boldsymbol{y}_n | X, \gamma_i) = \Big(\frac{1}{\sqrt{2\pi}\sigma_i}\Big)^{d}\exp(-\frac{||\boldsymbol{y}_{n}^{(i)} - \boldsymbol{y}_n||^2}{2\sigma_i^2}),
\end{equation}
$d$ is the output dimension, and $\boldsymbol{y}_{n}^{(i)}$ is the output from expert $i$ computed by the equations (\ref{equation:mixture_of_rnn_experts1}), (\ref{equation:mixture_of_rnn_experts2}) and (\ref{equation:mixture_of_rnn_experts3}) with parameter $\vartheta_i$.
Thus, the output of the model is governed by a mixture of normal distributions.
This equation results from the assumption that the observable sequence data is embedded in additive Gaussian noise.
It is well known that minimizing the mean square error is equivalent to maximizing the likelihood determined by a normal distribution for learning in a single neural network.
Therefore, equation (\ref{equation:mixture_of_normal_distribution}) is a natural extension of neural network learning.
The details of the derivation of equation (\ref{equation:mixture_of_normal_distribution}) are given in \cite{Jacobs91}.

Given a parameter set $\boldsymbol{\theta} = \big((\boldsymbol{\beta}_n)_{n=1}^{T}, \boldsymbol{\gamma}\big)$ and an input sequence $X$, the probability of an output sequence $Y = (\boldsymbol{y}_n)_{n=1}^T$ is given by
\begin{equation}
p(Y ~|~ X, \boldsymbol{\theta}) = \prod_{n=1}^{T} p(\boldsymbol{y}_n ~|~ X, \boldsymbol{\beta}_n, \boldsymbol{\gamma}).
\end{equation}
The likelihood function $L$ of a data set $D = (X, Y)$, parameterized by $\boldsymbol{\theta}$, is denoted by 
\begin{equation} \label{equation:likelihood_of_experts}
L(D, \boldsymbol{\theta}) = p(Y ~|~ X, \boldsymbol{\theta}) \varphi(\boldsymbol{\theta}),
\end{equation}
where $\varphi(\boldsymbol{\theta})$ is the p.d.f. of a prior distribution given by
\begin{equation} \label{equation:prior_distribution}
\varphi(\boldsymbol{\theta}) = \prod_{n=1}^{T-1} \prod_{i=1}^{N} \frac{1}{\sqrt{2\pi}\varsigma}\exp(-\frac{(\beta_{n+1}^{(i)} - \beta_n^{(i)})^2}{2\varsigma^2}).
\end{equation}
This equation means that the vector $\boldsymbol{\beta}_n$ is governed by $N$-dimensional Brownian motion.
The prior distribution has the effect of suppressing the change of gate opening values.

Assume that $D^{g} = (X, (\hat{\boldsymbol{g}}_n)_{n=1}^{T})$ is given.
We define the likelihood $L(D^{g}, \boldsymbol{\theta}^{g})$ by the Dirichlet distribution
\begin{equation}
p(\boldsymbol{g}_n ~|~ \hat{\boldsymbol{g}}_n) = \frac{1}{Z(\hat{\boldsymbol{g}}_n)}\prod_{i=1}^{N}\big(g_n^{(i)}\big)^{\hat{g}_n^{(i)}},
\end{equation}
\begin{equation} \label{equation:likelihood_of_gating_net}
L(D^{g}, \boldsymbol{\theta}^{g}) = \prod_{n=1}^{T} p(\boldsymbol{g}_n ~|~ \hat{\boldsymbol{g}}_n),
\end{equation}
where $Z(\hat{\boldsymbol{g}}_n)$ is the normalization constant, and $\boldsymbol{g}_n$ is given by $X$ and $\boldsymbol{\theta}^{g}$.
Notice that $\boldsymbol{g}_n$ in equation (\ref{equation:likelihood_of_gating_net}) is calculated by the teacher forcing technique \cite{Zipser89}, namely, using $\hat{\boldsymbol{g}}_{n-\tau}$ instead of $\boldsymbol{g}_{n-\tau}$ in equation (\ref{equation:mixture_of_rnn_experts5}).
In addition, because of the term $\boldsymbol{x}_{n-\tau}$ on the right-hand side of equation (\ref{equation:mixture_of_rnn_experts5}), $\boldsymbol{g}_n$ cannot be computed if $n \leq \tau$. 
Then it is assumed that $\boldsymbol{g}_n = \hat{\boldsymbol{g}}_n$ if $n \leq \tau$.
If we consider $g_n^{(i)}$ to be the probability of choosing an expert $i$ at time $n$, maximizing $\ln L(D^{g}, \boldsymbol{\theta}^{g})$ is equivalent to minimizing the Kullback-Leibler divergence
\begin{equation}
D_{\mathrm{KL}}((\hat{\boldsymbol{g}}_n)_{n=1}^{T} || (\boldsymbol{g}_n)_{n=1}^T) = \sum_{n=1}^{T}\sum_{i=1}^{N} \hat{g}_n^{(i)} \ln\Big(\frac{\hat{g}_n^{(i)}}{g_n^{(i)}}\Big).
\end{equation}

\subsubsection{Maximum likelihood estimation} \label{subsubsection:maximum_likelihood_estimation}

The learning method is to choose the best parameters $\boldsymbol{\theta}$ and $\boldsymbol{\theta}^{g}$ by maximizing (or integrating over) the likelihood $L(D, \boldsymbol{\theta})$ and $L(D^{g}, \boldsymbol{\theta}^{g})$.
More precisely, we use the gradient descent method with a momentum term as the training procedure.
The model parameters at learning step $t$ are updated according to 
\begin{equation} \label{equation:gradient_descent_method_with_momentum1}
\Delta\boldsymbol{\theta}^*(t) = \alpha\frac{\partial \ln L(D^*, \boldsymbol{\theta}^*(t))}{\partial \boldsymbol{\theta}^*(t)} + \eta\Delta\boldsymbol{\theta}^*(t-1) \qquad \mbox{where $(D^*, \boldsymbol{\theta}^*)$ is $(D, \boldsymbol{\theta})$ or $(D^{g}, \boldsymbol{\theta}^{g})$},
\end{equation}
\begin{equation} \label{equation:gradient_descent_method_with_momentum2}
\boldsymbol{\theta}^*(t+1) = \boldsymbol{\theta}^*(t) + \Delta\boldsymbol{\theta}^*(t),
\end{equation}
where $\alpha$ is the learning rate and $\eta$ is a momentum term parameter.
For each parameter, the partial differential equations $\frac{\partial \ln L(D, \boldsymbol{\theta})}{\partial \boldsymbol{\theta}}$ are given by
\begin{equation} \label{equation:gradient_beta}
\frac{\partial \ln L(D, \boldsymbol{\theta})}{\partial \beta_{n}^{(i)}} = \frac{\hat{g}_{n}^{(i)} (p(\boldsymbol{y}_n ~|~ X, \gamma_i) -  p(\boldsymbol{y}_n ~|~ X, \boldsymbol{\beta}_n, \boldsymbol{\gamma}))}{p(\boldsymbol{y}_n ~|~ X, \boldsymbol{\beta}_n, \boldsymbol{\gamma})} + \frac{G_{n}^{(i)}}{\varsigma^2},
\end{equation}
\begin{equation}
G_{n}^{(i)} = {\scriptsize \left\{ \begin{array}{cl}
\beta_{n+1}^{(i)} - \beta_{n}^{(i)} & \mbox{if $n = 1$}, \\
-\beta_{n}^{(i)} + \beta_{n-1}^{(i)} & \mbox{if $n = T$}, \\
\beta_{n+1}^{(i)} - 2\beta_{n}^{(i)} + \beta_{n-1}^{(i)} & \mbox{otherwise},
\end{array} \right.}
\end{equation}
\begin{equation} \label{equation:gradient_vartheta}
\frac{\partial \ln L(D, \boldsymbol{\theta})}{\partial \vartheta_i} = \sum_{n=1}^{T} \frac{\hat{g}_{n}^{(i)} p(\boldsymbol{y}_n ~|~ X, \gamma_i)}{p(\boldsymbol{y}_n ~|~ X, \boldsymbol{\beta}_n, \boldsymbol{\gamma})} \Big[ \frac{-1}{2\sigma_i^2}\frac{\partial}{\partial \vartheta_i}||\boldsymbol{y}_{n}^{(i)} - \boldsymbol{y}_n||^2 \Big],
\end{equation}
\begin{equation} \label{equation:gradient_sigma}
\frac{\partial \ln L(D, \boldsymbol{\theta})}{\partial \sigma_i} = \sum_{n=1}^{T} \frac{\hat{g}_{n}^{(i)} p(\boldsymbol{y}_n ~|~ X, \gamma_i)}{p(\boldsymbol{y}_n ~|~ X, \boldsymbol{\beta}_n, \boldsymbol{\gamma})} \Big[ -\frac{d}{\sigma_i} + \frac{||\boldsymbol{y}_{n}^{(i)} - \boldsymbol{y}_n||^2}{\sigma_i^3} \Big],
\end{equation}
and $\frac{\partial \ln L(D^{g}, \boldsymbol{\theta}^{g})}{\partial \boldsymbol{\theta}^{g}}$ is also given by
\begin{equation}
\frac{\partial \ln L(D^{g}, \boldsymbol{\theta}^{g})}{\partial \boldsymbol{\theta}^{g}} = \sum_{n=1}^{T} \sum_{i=1}^{N}\frac{\hat{g}_n^{(i)}}{g_n^{(i)}} \big( \frac{\partial g_n^{(i)}}{\partial \boldsymbol{\theta}^{g}} \big).
\end{equation}
Note that $\frac{\partial}{\partial \vartheta_i}||\boldsymbol{y}_{n}^{(i)} - \boldsymbol{y}_n||^2$ and $\frac{\partial g_n^{(i)}}{\partial \boldsymbol{\theta}^{g}}$ can be solved for by the back propagation through time (BPTT) method \cite{Rumelhart86}.
In this paper, we assume that the infimum $\bar{\sigma}$ of the parameter $\sigma_i$ is greater than $0$, because if $\sigma_i$ converges to $0$, then $\Delta\theta(t)$ diverges.

We have explained the case in which the training data set $D$ is a single sequence, but the method can be easily extended to learning of several sequences by using the sum of the gradients for each sequence.
When several sequences are used as training data, initial states $\boldsymbol{u}_0^{(i)}$ and $\boldsymbol{u}_0^g$, and an estimate $\hat{\boldsymbol{g}}_n$ of the gate opening vector have to be provided for each sequence.

\subsubsection{Acceleration of gating network learning} \label{subsubsection:acceleration_of_the_gating_network_learning}

In principle, our training procedure is defined by equations (\ref{equation:gradient_descent_method_with_momentum1}) and (\ref{equation:gradient_descent_method_with_momentum2}).
However, for reasons which will be explained in section \ref{subsection:problem_in_learning_stochastic_time_series_as_deterministic_dynamics_of_chaos}, learning for a gating network sometimes becomes unstable, and so the likelihood $L(D^{g}, \boldsymbol{\theta}^{g})$ often decreases when updating the parameter $\boldsymbol{\theta}^{g}$ with certain learning rates.
Of course, if we use sufficiently small learning rates, the likelihood does not decrease for any learning steps; however, it does require considerable computation time.
Hence, in order to accelerate gating network learning practically, we update the learning rate adaptively by means of the following algorithm.
\begin{ol}{2}
\item For each learning step, updated parameters are computed by equations (\ref{equation:gradient_descent_method_with_momentum1}) and (\ref{equation:gradient_descent_method_with_momentum2}) applying the learning rate $\alpha$, and the rate $r$ defined by
\begin{equation}
r = \frac{D_{\mathrm{KL}}((\hat{\boldsymbol{g}}_n)_{n=1}^{T} || (\boldsymbol{g}'_n)_{n=1}^T)}{D_{\mathrm{KL}}((\hat{\boldsymbol{g}}_n)_{n=1}^{T} || (\boldsymbol{g}_n)_{n=1}^T)}
\end{equation}
is also computed, where $(\boldsymbol{g}_n)_{n=1}^T$ and $(\boldsymbol{g}'_n)_{n=1}^T$ are sequences of gate opening values corresponding to the current parameters and the updated parameters, respectively.
\item If $r > r_{\mathrm{th}}$, then $\alpha$ is replaced with $\alpha\alpha_{\mathrm{dec}}$ and we return to (1).
Otherwise go to (3).
\item If $r < 1$, then $\alpha$ is replaced with $\alpha \alpha_{\mathrm{inc}}$.
Go to the next learning step.
\end{ol}
In this study, we use $r_{\mathrm{th}} = 1.1$, $\alpha_{\mathrm{dec}} = 0.7$ and $\alpha_{\mathrm{inc}} = 1.05$.

\subsubsection{Initialization} \label{subsubsection:initialization}

Here, we describe a policy for initialization of parameters and setting a learning rate in this paper.
Every element of matrices and thresholds of a network, either an expert or a gating network, is initialized by choosing randomly from the uniform distribution on the interval $[-\frac{1}{M},\frac{1}{M}]$, where $M$ is the number of context neurons.
Initial states $\boldsymbol{u}_0^{(i)}$ and $\boldsymbol{u}_0^g$ are also initialized randomly from the interval $[-1,1]$.
The parameters $(\boldsymbol{\beta}_n)_{n=1}^T$ and $\boldsymbol{\sigma}$ are initialized such that $\boldsymbol{\beta}_n = \boldsymbol{0}$ for each $n$ and $\boldsymbol{\sigma} = \boldsymbol{1}$, respectively.
Since the maximum value of $L(D, \boldsymbol{\theta})$ depends on the total length $T$ of training sequences and the dimension $d$ of output neurons, we scale the learning rate $\alpha$ by a parameter $\tilde{\alpha}$ satisfying $\alpha = \frac{1}{Td}\tilde{\alpha}$.

\subsection{Feedback loop with time delay} \label{subsection:open_and_closed_loop}

Let us consider the case in which there exists a feedback loop from output to input with a time delay $\tau$, that is to say, the model is an autonomous system.
In this case, a training data set $D = (X, Y)$ has to satisfy $\boldsymbol{y}_n = \boldsymbol{x}_{n+\tau}$.
At the end of learning, if every output of the model is completely equal to the training data, the model can generate the sequence $Y$ using a feedback loop without external input data $X$.
In this paper, the dynamics of the model with external input is referred to as open-loop dynamics, and dynamics with self-feedback is also called closed-loop dynamics.
In section \ref{section:simulation}, we consider the case in which the model is an autonomous system.

\section{Numerical simulation} \label{section:simulation}

We now demonstrate how the mixture of RNN experts learns to generate sequences given as tasks.
Our first experiment in section \ref{subsection:learning_stochastic_finite_state_automaton} is to learn a stochastic finite state automaton by means of a gating network, in order to test whether a gating network can output an estimate of the gate opening vector even if switching of winners is stochastic.
We include this test for two reasons.
(1) Since a time series of gate opening vectors represents an arbitrary composition of primitives, such a time series is usually a symbolic and nondeterministic sequence.
(2) It has been shown that the simple recurrent network (SRN or Elman net) can learn symbolic and nondeterministic sequences such as declarative sentences \cite{Elman90,Elman91}.
The SRN has typically been used to learn to predict the probability of occurrence of the next symbol, and the output values of the SRN represented this probability.
For example, if each symbol is to appear next with equal probability, the output values of the SRN are constant, summing to one.
However, in the context of the mixture of RNN experts model, the weighted sum of output values of all experts by probability does not represent a correct value.
In order to output the correct value, the gating network has to determine a winner among the experts, and the output of the gating network corresponding to the winner should be almost one.
This means that the gating network has to learn to output not the probability but stochastic variables instead.
It is not trivial to determine whether the gating network can learn this task, because the network is a deterministic dynamical system.
At least, if the maximum Lyapunov exponent of the network is a negative value, the output of the network is periodic and so cannot represent any nondeterministic sequences.
Basically, a necessary condition to output nondeterministic sequences using a deterministic dynamical system is that the maximum Lyapunov exponent is positive.

On the other hand, it is known that there are chaotic dynamics which resemble nondeterministic sequences, because it is possible to create a correspondence between a hyperbolic dynamical system and a symbolic dynamics by constructing a Markov partition \cite{Lind95}.
Tani et al. \cite{NishimotoNN04,Tani95b} demonstrated that a nondeterministic symbolic process can be imitated by an RNN trained by the gradient descent method, utilizing initial sensitivity characteristics of a chaotic dynamics.
However, they only investigated the case in which the length of training data is about $10$.
Thus, in our first experiment we show that the gating network can acquire a chaotic dynamics and can generate nondeterministic sequences, even if training data are much longer.

Our second experiment in section \ref{subsection:learning_stochastically_switching_of_Lissajous_curves} is to learn Markov chains whose primitives are Lissajous curve patterns.
It was shown that a mixture of RNN experts can extract a set of reusable primitives from training data generated by a Markov chain of Lissajous curves \cite{Namikawa2008}, but learning for the gating network was not considered.
Therefore, we will show that a mixture of RNN experts can generate stochastically combined Lissajous curves by switching winners using a gating network.

In order to evaluate the learning capability of a mixture of RNN experts, we first define some measures of fitness.
An error $E$ for expert networks is defined by the mean square error
\begin{equation}
E = \frac{1}{2Td} \sum_{n=1}^{T} ||\boldsymbol{{y}}_n^{\mathrm{teach}} - \boldsymbol{\hat{y}}_{n}||^2
\end{equation}
for each learning step, where $\boldsymbol{y}_n^{\mathrm{teach}}$ denotes the training data, and $\boldsymbol{\hat{y}}_{n} = \sum_{i=1}^{N}\hat{g}_n^{(i)}\boldsymbol{y}_n^{(i)}$.
An error $E^{g}$ for the gating network is defined by the Kullback-Leibler divergence
\begin{eqnarray}
E^{g} & = & \frac{1}{TN} D_{\mathrm{KL}}((\hat{\boldsymbol{g}}_n)_{n=1}^{T} || (\boldsymbol{g}_n)_{n=1}^T) \nonumber \\
& = & \frac{1}{TN} \sum_{n=1}^{T}\sum_{i=1}^{N} \hat{g}_n^{(i)} \ln\Big(\frac{\hat{g}_n^{(i)}}{g_n^{(i)}}\Big),
\end{eqnarray}
where we assume that $E$ and $E^{g}$ are computed by teacher forcing, i.e., utilizing the training data and estimates of gate opening values as network input.
If $E = E^{g} = 0$, then the mixture of RNN experts can output sequences equivalent to the training data.
Since $E$ and $E^{g}$ are measures for open-loop dynamics, we next define the Kullback-Leibler divergence to evaluate fitness in the closed-loop dynamics.
\begin{equation}
D_{\mathrm{KL}}^{(m)}((\hat{s}_n)_{n=1}^{T} || (s_n)_{n=1}^T) = \sum_{\boldsymbol{s} \in \{1,\cdots,N\}^m} p(\boldsymbol{s}|(\hat{s}_n)_{n=1}^{T}) \ln\frac{p(\boldsymbol{s}|(\hat{s}_n)_{n=1}^{T})}{p(\boldsymbol{s}|(s_n)_{n=1}^T)},
\end{equation}
where $s_n$ denotes the index of a current opening gate $s_n \in \{1,\cdots,N\}$ defined by
\begin{equation} \label{equation:index_of_dominant_network}
s_n = \arg\max_i g_n^{(i)}
\end{equation}
and $\hat{s}_n$ is given by equation (\ref{equation:index_of_dominant_network}) with respect to $\hat{\boldsymbol{g}}_n$.
By using this measure, we can perform an evaluation even if the training data are stochastic time series combining several primitive patterns.

\subsection{Preliminary experiment: Learning a stochastic finite state automaton} \label{subsection:learning_stochastic_finite_state_automaton}

\begin{figure}
\begin{center}
\scalebox{1.0}{\includegraphics{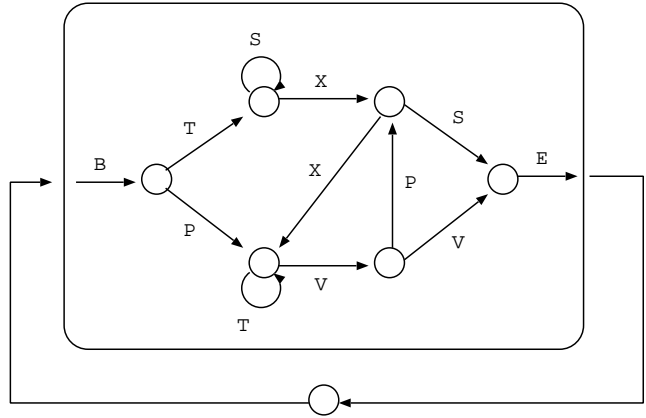}}
\caption{Finite state automaton based on the Reber grammar.
If there are two paths, then we randomly choose one with equal probability.
}
\label{figure:Reber_grammar}
\end{center}
\end{figure}

Here, we investigate how a gating network learns time series of gate opening values which form stochastic sequences.
As we focus on training of the gating network, we assume that the dimension of the input $\boldsymbol{x}_n$ is $d = 0$.
In other words, any effects of experts and the input-output sequence $D = (X,Y)$ are omitted.
As a benchmark system, we use a finite state automaton described in Figure \ref{figure:Reber_grammar}.
This automaton is given by modifying the Reber grammar to an Ergodic model.
When we generate a sequence using the automaton, we choose a path randomly from two paths with equal probability.
Representation of the symbols by gate opening values is such that each symbol is mapped with a one to one correspondence to an integer from $1$ to $7$, and if a symbol corresponding to $i$ appears at time $n$, then $\hat{g}_n^{(i)} = 1$.

\begin{figure}
\begin{center}
\scalebox{1.0}{\includegraphics{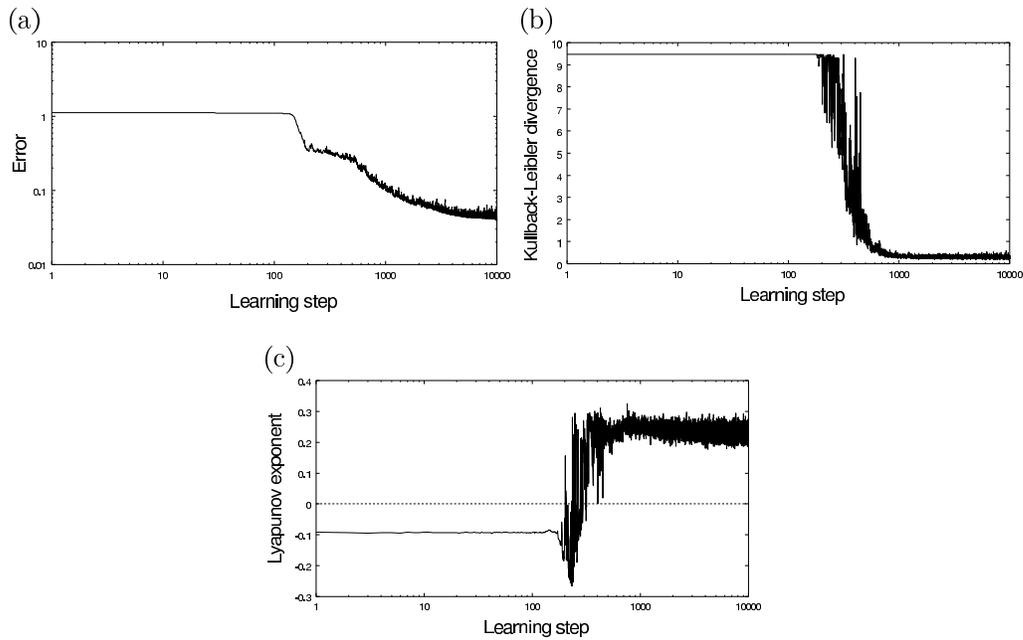}}
\caption{(a) The error $E^g$.
(b) Kullback-Leibler divergence $D_{\mathrm{KL}}^{(3)}((\hat{s}_n)_{n=1}^{T} || (s_n)_{n=1}^T)$.
(c) Maximum Lyapunov exponent.}
\label{figure:error_kldiv_lyapunov_for_reber_grammar}
\end{center}
\end{figure}

\begin{figure}
\begin{center}
\scalebox{1.0}{\includegraphics{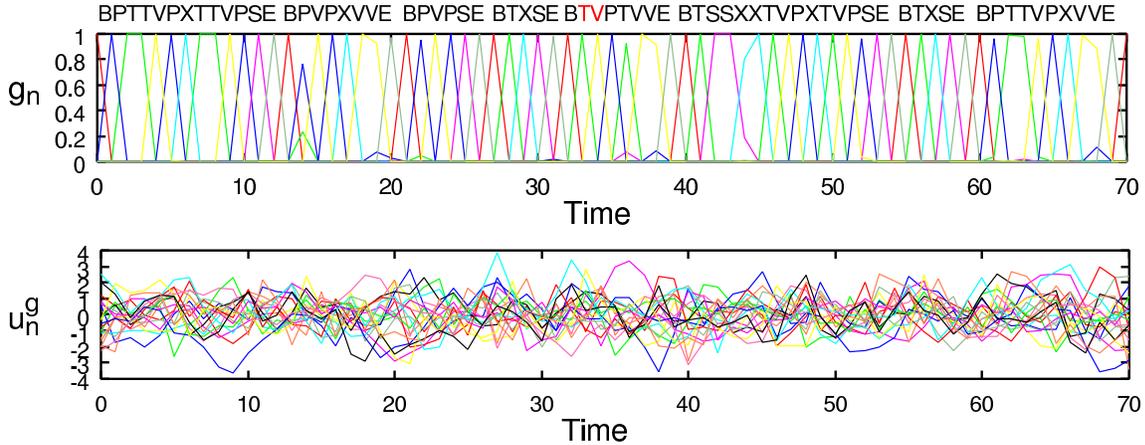}}
\caption{A snapshot of $\boldsymbol{g}_n$ and $\boldsymbol{u}_{n}^{g}$ of the trained network with closed-loop dynamics.
Each label over gate opening values denotes a symbol corresponding to the current opening gate, and a red label means that the symbol does not satisfy the syntax of the Reber grammar.
}
\label{figure:reber_grammar_rnn_orbit}
\end{center}
\end{figure}

As an example, we computed learning for the gating network up to $10000$ steps.
Here we use $4$ training sequences of length $T = 500$ (so the total length of training sequences is $2000$).
The number of context neurons, time constant, time delay of a feedback loop, learning rate and momentum were chosen to be $20$, $\epsilon^g = 0.1$, $\tau = 1$, $\tilde{\alpha} = 0.01$ and $\eta = 0.9$, respectively.
In Figure \ref{figure:error_kldiv_lyapunov_for_reber_grammar}, we depict the error, Kullback-Leibler divergence and maximum Lyapunov exponent for each learning step, where the Lyapunov exponent is computed for the closed-loop dynamics.
Here, we computed the Kullback-Leibler divergence with $m = 3$, because the stochastic process defined by Figure \ref{figure:Reber_grammar} is a second order Markov chain.
Since the maximum Lyapunov exponent becomes positive, the gating network acquires a chaotic dynamics by training.
Moreover, acquiring the chaotic dynamics is effective for imitation of the stochastic process, because the Kullback-Leibler divergence decreases with the increase in the maximum Lyapunov exponent.
Figure \ref{figure:reber_grammar_rnn_orbit} displays the dynamics of the trained network with a random initial state.
It can be seen that the output of the network is almost always $0$ or $1$.
This result indicates that the network learned to output not the probability but stochastic variables.
Furthermore, there is a case for which the output of the network does not satisfy the syntax of the Reber grammar (printed as red color symbols over the gate opening values in Figure \ref{figure:reber_grammar_rnn_orbit}), but the dynamics recovers the correct output patterns.

Table \ref{table:result_of_reber_grammar} shows the results of training for each number of context neurons.
For each case, we computed $50$ samples up to $10000$ learning steps with different initial conditions, where we use the same parameters as for the above training example without the number of context neurons.
In addition, in order to evaluate the generalization capability of trained networks, we prepared test data separate from the training data, and we computed the Kullback-Leibler divergence $D_{\mathrm{KL}}^{(m)}((\hat{s}_n)_{n=1}^{T} || (s_n)_{n=1}^T)$ for the test data with random initial states.
It can be seen that if there exist sufficient context neurons to learn the training data, the dynamics of some trained networks become chaotic and the Kullback-Leibler divergences between sequences generated by the networks and training/test data become sufficiently small.

\begin{table}
\caption{The results of training (mean of $50$ trials).
A percentage in round brackets represents the ratio of trained networks whose maximum Lyapunov exponent is positive.
When we evaluated a Lyapunov exponent, we computed $200$ sample sequences of $10,000$ time steps for each trained network.}
\label{table:result_of_reber_grammar}
\begin{center}
\begin{tabular}{ccccc}
\hline 
Number of & Error & \multicolumn{2}{c}{{\small Kullback-Leibler divergence} {\scriptsize $D_{\mathrm{KL}}^{(m)}((\hat{s}_n)_{n=1}^{T} || (s_n)_{n=1}^T)$}} & Lyapunov \\
context neurons & $E^g$ & (for training data) & (for test data) & exponent \\
\hline
10 & 0.144868 & 2.511791 & 2.388665 & 0.203512 (88\%) \\
20 & 0.010118 & 0.258037 & 0.309089 & 0.215117 (98\%) \\
30 & 0.005721 & 0.051233 & 0.327273 & 0.160728 (100\%) \\
50 & 0.004339 & 0.006338 & 0.290858 & 0.116574(100\%) \\
\hline
\end{tabular}
\end{center}
\end{table}

\subsection{Learning stochastic switching of Lissajous curves} \label{subsection:learning_stochastically_switching_of_Lissajous_curves}

In the current section, we consider two tasks, which are to generate time series stochastically combining Lissajous curves.
Assume that for each task, transitions among curves are consonant with orbit continuity.
Since we consider a model which has a feedback loop with time delay $\tau$, every training data set $D = (X, Y)$ satisfies $\boldsymbol{y}_n = \boldsymbol{x}_{n+\tau}$.

\subsubsection{Two Lissajous curves case} \label{subsubsection:two_Lissajous_curves_case}

\begin{figure}
\begin{center}
\scalebox{1.0}{\includegraphics{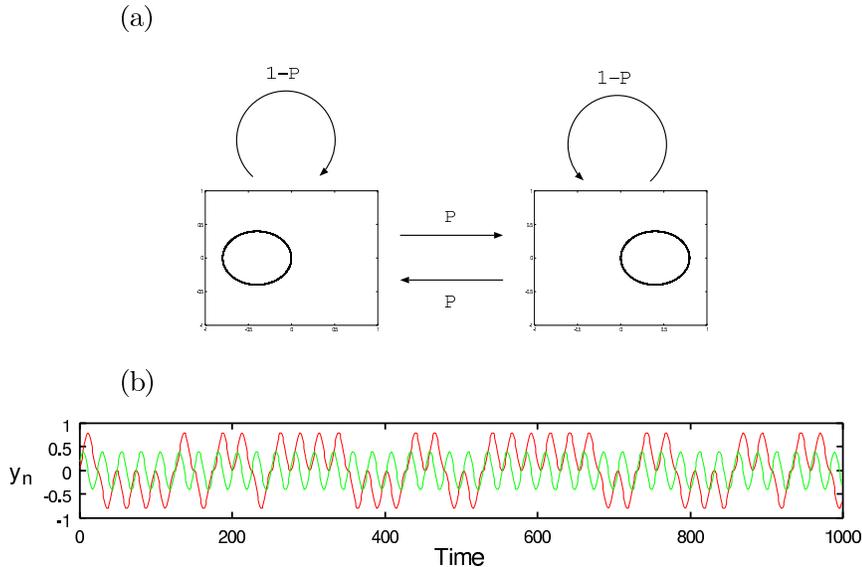}}
\caption{(a) Transition rule of the training data.
Each figure represents the trajectory of a Lissajous curve, and $P$ denotes the transition probability between curves.
(b) An example of the training data.}
\label{figure:transition_rule_and_teaching_data}
\end{center}
\end{figure}

The first task is to learn $2$-dimensional sequences generated by Markov chain switching of $2$ Lissajous curves of period $25$ (see Figure \ref{figure:transition_rule_and_teaching_data}).
We use $10$ training sequences each of length $T = 1000$ and with transition probability $P = 0.5$.
The time delay of the feedback loop is $\tau = 1$.
The number of experts is $N = 2$.
There are $10$ context neurons for each expert and $15$ for the gating network.
The parameter settings are $\epsilon = 0.2$, $\epsilon^g = 0.04$, $\bar{\sigma} = 0.05$, $\varsigma = 10$, $\tilde{\alpha} = 0.01$ and $\eta = 0.9$.

\begin{figure}
\begin{center}
\scalebox{1.0}{\includegraphics{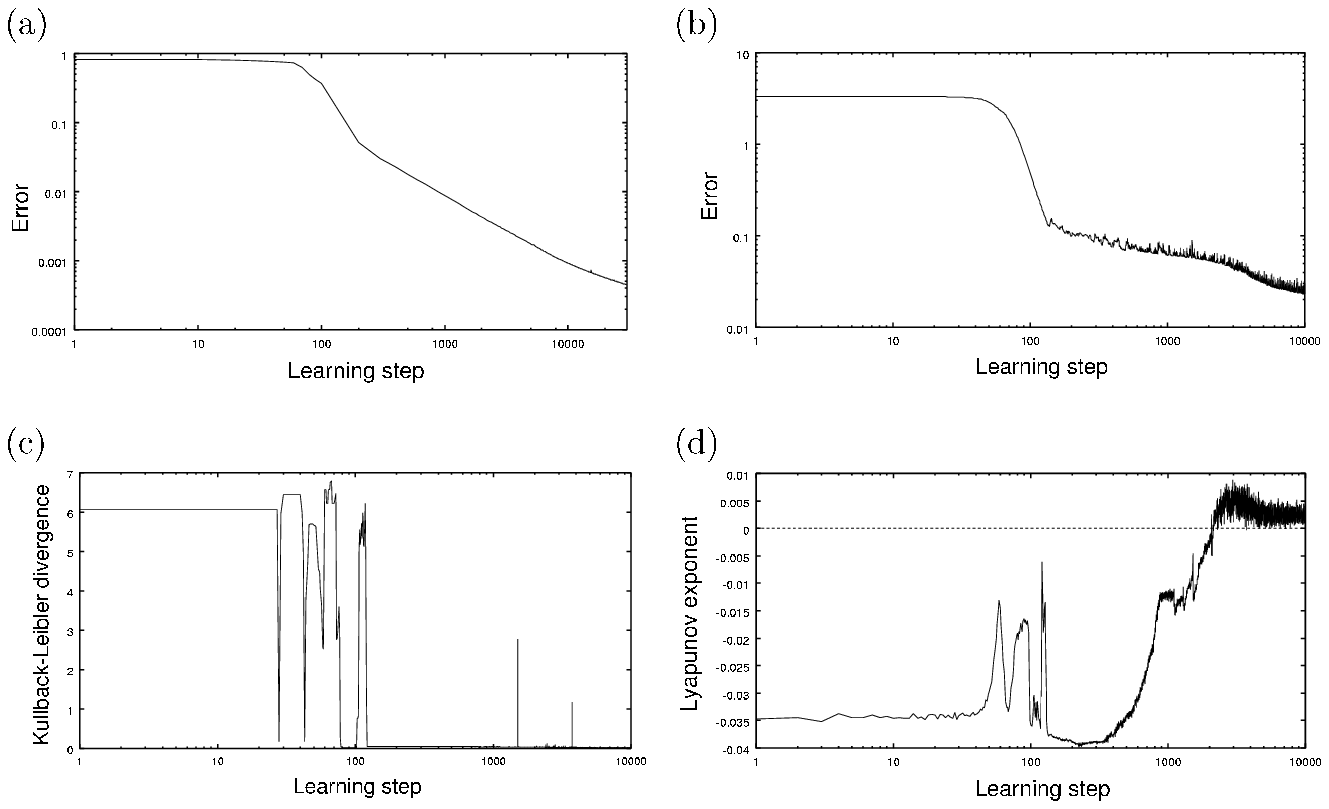}}
\caption{(a) The error of experts $E$.
(b) The error of the gating network $E^g$.
(c) The Kullback-Leibler divergence $D_{\mathrm{KL}}^{(5)}((\hat{s}_n)_{n=1}^{T} || (s_n)_{n=1}^T)$.
(d) Maximum Lyapunov exponent.}
\label{figure:error_kldiv_lyapunov_for_Lissajous2}
\end{center}
\end{figure}

\begin{figure}
\begin{center}
\scalebox{1.0}{\includegraphics{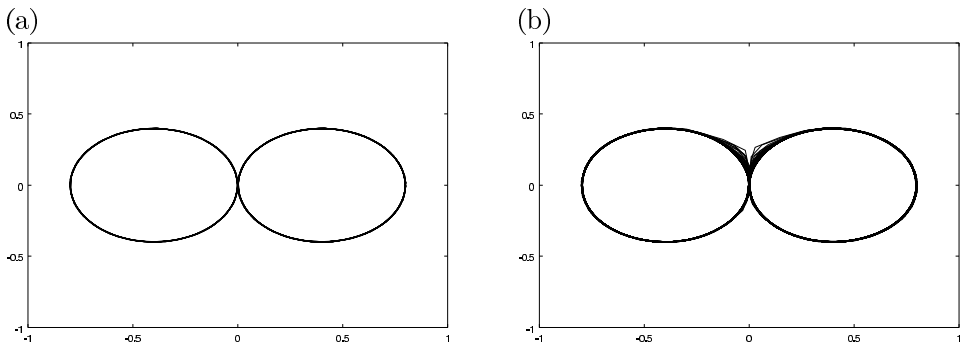}}
\caption{(a) Teaching data.
(b) A trajectory of the mixture of RNN experts with closed-loop dynamics (plotted over $10,000$ time steps after convergence of the transients).}
\label{figure:2d_teach_and_output_for_Lissajous2}
\end{center}
\end{figure}

\begin{figure}
\begin{center}
\scalebox{1.0}{\includegraphics{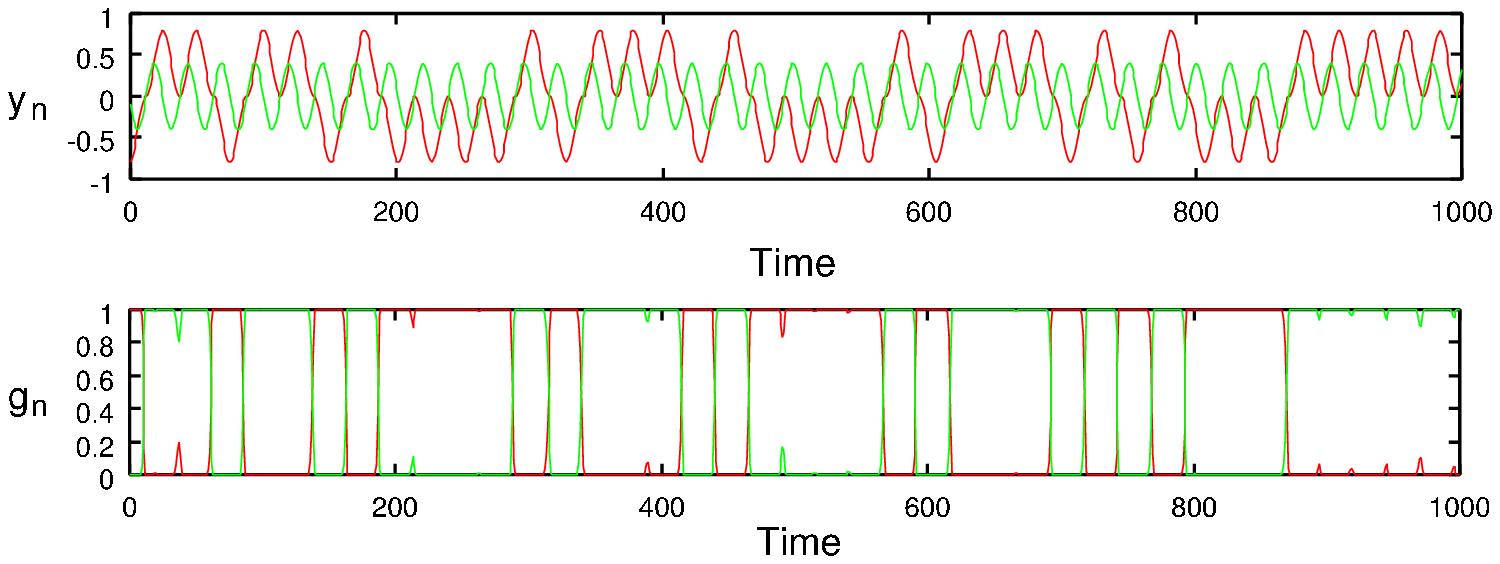}}
\caption{Output $\boldsymbol{y}_n$ and gate opening vector $\boldsymbol{g}_{n}$ of the mixture of RNN experts with closed-loop dynamics.}
\label{figure:orbit_for_Lissajous2}
\end{center}
\end{figure}

Figure \ref{figure:error_kldiv_lyapunov_for_Lissajous2} depicts an example of results for training the mixture of RNN experts with this task.
Here, we computed learning for experts up to $30000$ steps, and learning for the gating network up to $10000$ steps.
Figure \ref{figure:2d_teach_and_output_for_Lissajous2} and \ref{figure:orbit_for_Lissajous2} depict trajectories of the trained network computed from the closed-loop dynamics.
Figure \ref{figure:2d_teach_and_output_for_Lissajous2} indicates that the trained network can generate trajectories similar to the training data.
Moreover, from Figure \ref{figure:error_kldiv_lyapunov_for_Lissajous2} (d) and Figure \ref{figure:orbit_for_Lissajous2}, it can be observed that the network has acquired stochastic switching between Lissajous curves using a chaotic dynamics.

\begin{figure}
\begin{center}
\scalebox{0.9}{\includegraphics{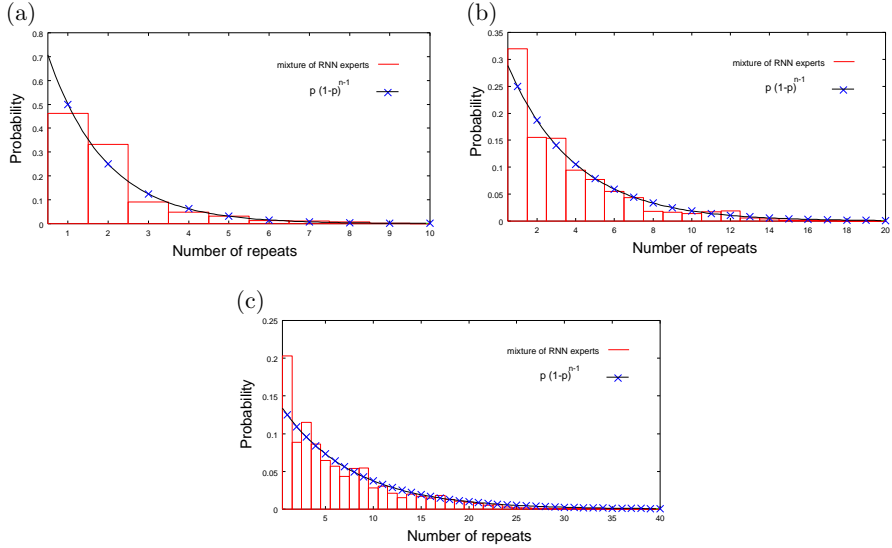}}
\caption{Distributions of the number of repetitions of each Lissajous curve, for each transition probability $P$ of training data.
(a) $P = 0.5$,
(b) $P = 0.25$,
(c) $P = 0.125$.
Here, a histogram describes an averaged distribution of a whole set of trained networks, and a line describes the theoretical distribution of training data.
When we evaluated the distributions, we computed $50$ sample sequences of $100,000$ time steps for each network.
}
\label{figure:residence_time_distribution_for_Lissajous2}
\end{center}
\end{figure}

In Table \ref{table:result_of_Lissajous2} and Figure \ref{figure:residence_time_distribution_for_Lissajous2}, to show that the model can acquire probabilities of transition, we compare cases in which the training data are generated by different transition probabilities.
Here, we present evaluated results of $30$ samples for each case with the same parameters as for the previous training.
Table \ref{table:result_of_Lissajous2} describes transition probabilities between curves and maximum Lyapunov exponents, and Figure \ref{figure:residence_time_distribution_for_Lissajous2} also describes probability distributions for the number of repetitions of each curve.
These results confirm that the mixture of RNN experts can acquire the transition probabilities of the task.

\begin{table}
\caption{The results of training (mean of $30$ trials).
Here, a probability refers to a transition probability between curves.
When we evaluated the probability and the Lyapunov exponent, we computed $50$ sample sequences of $100,000$ time steps for each network.}
\label{table:result_of_Lissajous2}
\begin{center}
\begin{tabular}{ccc}
\hline 
Probability & Probability     & Lyapunov \\
(task)      & (trained network) & exponent \\
\hline
0.5   & 0.451033 & 0.001076 (93\%) \\
0.25  & 0.275999 & 0.002080 (96\%) \\
0.125 & 0.150866 & 0.002829 (96\%) \\
\hline
\end{tabular}
\end{center}
\end{table}

\subsubsection{Multiple Lissajous curves case} \label{subsubsection:multiple_Lissajous_curves_case}

\begin{figure}
\begin{center}
\scalebox{1.0}{\includegraphics{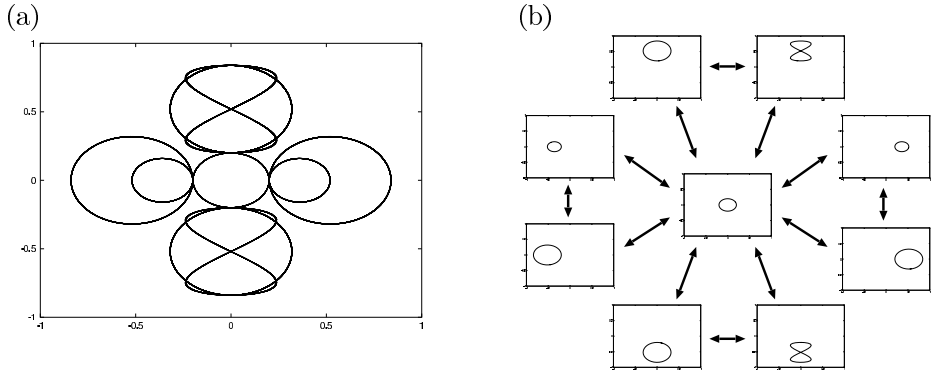}}
\caption{(a) Training data generated by switching among $9$ Lissajous curves.
(b) Each Lissajous curve.
Transition among curves is consonant with continuity of the orbit.
Keeping a current curve or moving to another curve is chosen randomly with equal probability.
}
\label{figure:teaching_data_for_Lissajous9}
\end{center}
\end{figure}

The second task is to learn time series defined by switching among $9$ Lissajous curves of period $32$ (see Figure \ref{figure:teaching_data_for_Lissajous9}).
In the task, we use a training sequence of length $T = 10000$ with a time delay of feedback $\tau = 5$.
We have $N = 24$ experts, each with $10$ context neurons, and the gating network has $30$ context neurons.

\begin{figure}
\begin{center}
\scalebox{1.0}{\includegraphics{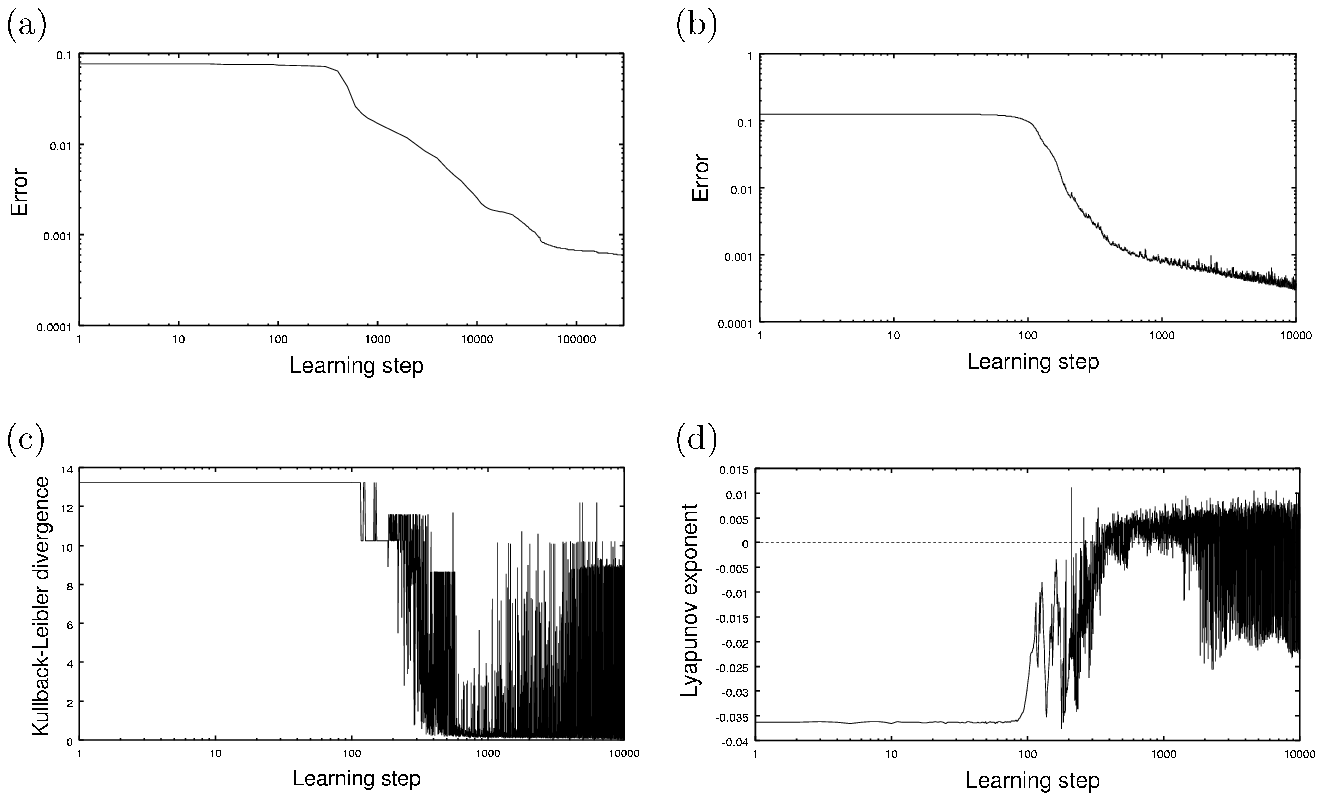}}
\caption{(a) The error of the experts $E$.
(b) The error of the gating network $E^g$.
(c) Kullback-Leibler divergence $D_{\mathrm{KL}}^{(5)}((\hat{s}_n)_{n=1}^{T} || (s_n)_{n=1}^T)$.
(d) Maximum Lyapunov exponent.}
\label{figure:error_kldiv_lyapunov_for_Lissajous9}
\end{center}
\end{figure}

In Figure \ref{figure:error_kldiv_lyapunov_for_Lissajous9}, we display an example of training, where parameters were set as follows: $\epsilon = 0.1$, $\epsilon^g = 0.04$, $\bar{\sigma} = 0.05$, $\varsigma = 1$, $\tilde{\alpha} = 0.01$ and $\eta = 0.9$.
It can be seen that the maximum Lyapunov exponent oscillates between positive and negative values after $1000$ learning steps; concurrently the Kullback-Leibler divergence changes drastically.
This phenomenon implies that properties of the network as a dynamical system are changed by updating parameters, and the network is not structurally stable.
In the case, the stability of varying a likelihood (or an error) is not corresponding to that of varying the properties of the model.
Actually, in Figure \ref{figure:error_kldiv_lyapunov_for_Lissajous9} the error decreases monotonically, whereas the maximum Lyapunov exponent oscillates.
In general, since the topological property of a dynamical system being not structurally stable is changed by perturbations \cite{Andronov1937}, it is difficult to prevent oscillation of the maximum Lyapunov exponent even when the learning rate is very small.
In section \ref{subsection:structural_stability_of_learning_models}, we will explain and discuss this problem of the structural stability of learning models in detail.

\begin{figure}
\begin{center}
\scalebox{1.0}{\includegraphics{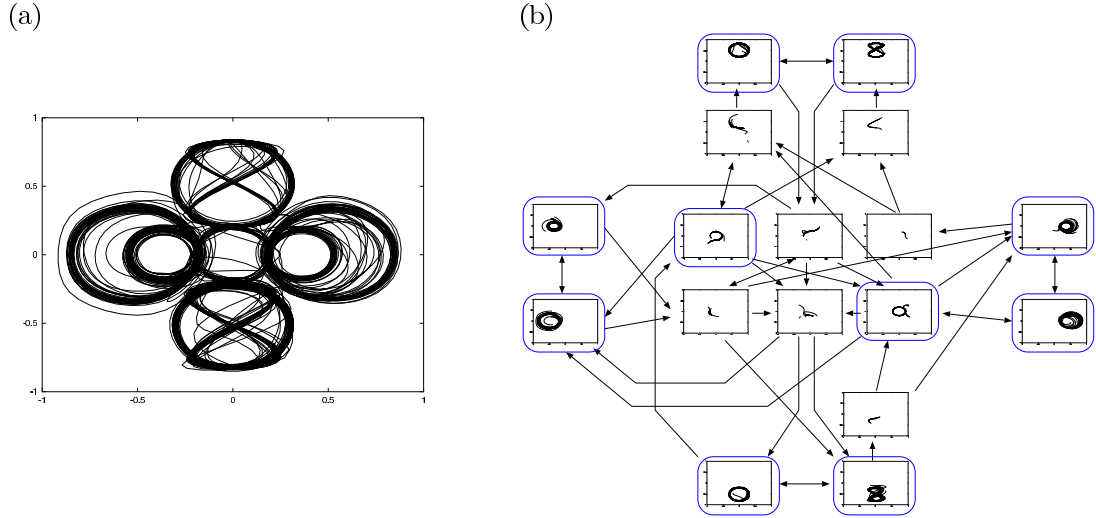}}
\caption{(a) A trajectory generated by the trained network in the closed-loop dynamics (plotted over $10,000$ time steps after convergence of the transients).
(b) Output of experts and transition paths.
The output of an expert $i$ is plotted if $s_n = i$, namely, if gate $i$ opens at time $n$.
If gate $i$ is never opened, then no output of the expert $i$ is plotted.
The graphs enclosed by blue boxes represent Lissajous curves as primitives, and other graphs represent paths connecting curves.
}
\label{figure:output_for_Lissajous9}
\end{center}
\end{figure}

\begin{figure}
\begin{center}
\scalebox{0.8}{\includegraphics{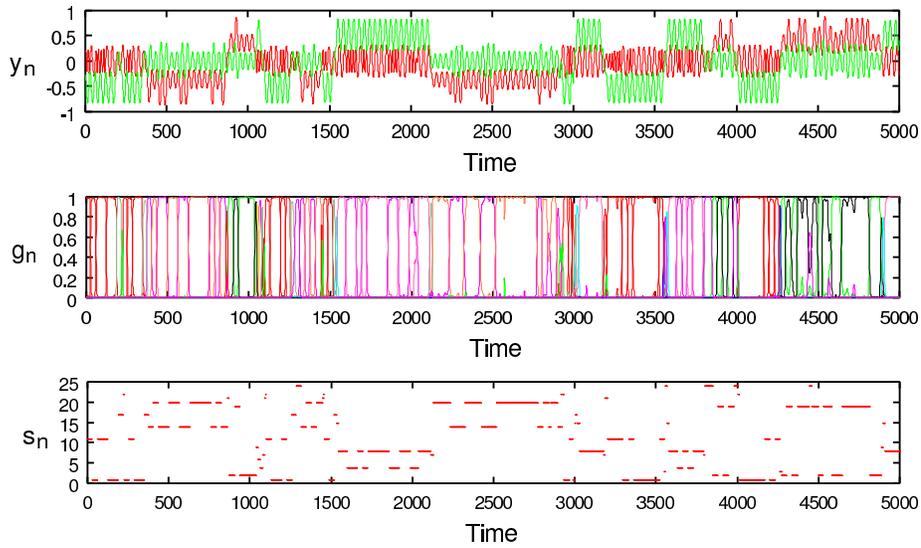}}
\caption{A snapshot of output $\boldsymbol{y}_n$, gate opening vector $\boldsymbol{g}_{n}$ and index of a current opening gate $s_n$ of the trained network.}
\label{figure:orbit_for_Lissajous9}
\end{center}
\end{figure}

Figures \ref{figure:output_for_Lissajous9} and \ref{figure:orbit_for_Lissajous9} represent closed-loop dynamics of the trained network with a random initial state, where the network has a positive Lyapunov exponent.
It can be seen in Figure \ref{figure:output_for_Lissajous9} (b) that there are two groups of experts; one group represents primitives, and the other represents transition paths between primitives.
Furthermore, taking into account the whole set of transition paths, the transition rule for the network described in Figure \ref{figure:output_for_Lissajous9} (b) is similar to that for the training data described in Figure \ref{figure:teaching_data_for_Lissajous9} (b).

\subsection{Summary of results} \label{subsection:summary_of_results}

In the preliminary experiment, we tested the gating network for its capability to learn stochastic sequences given by a finite state automaton with a chaotic dynamics.
The result shows that, by increasing the number of context neurons, the Kullback-Leibler divergence can be reduced and the frequency of appearance of chaos also increases.

For the next experiment, we considered the task of learning stochastic time series consisting of Lissajous curves.
In the case of two Lissajous curves, a mixture of RNN experts can imitate by chaos not only nondeterministic characteristics but a probability distribution underlying training data.
Moreover, the networks can maintain a chaotic dynamics stably at the end of learning.
In the case of multiple Lissajous curves, the trained network can generate a similar sequence to the training data, though the maximum Lyapunov exponent is not always positive in the last learning phase.
Since this instability characteristic of chaos is due to the structurally unstable dynamics of the network, we will discuss this problem of structural stability in section \ref{subsection:structural_stability_of_learning_models}.

\section{Analysis and Discussion} \label{section:discussion}

In this paper, we have investigated learning processes of a gating network contained in the mixture of RNN experts model.
In particular, we have focused on how the gating network learns stochastic sequences brought by an arbitrary composition of primitives by self-organizing chaos.
The fact that the gating network can learn stochastic time series is important, because it had been considered that it is impossible for RNNs of deterministic dynamical systems to learn nondeterministic time series.

In the following we discuss some general characteristics in the learning of stochastic time series by embedding into chaos.
We focus on two essential problems, related to memory capacity and to the structural stability of RNNs, and explain how the problems can be overcome.

\subsection{A problem in learning stochastic time series by means of a deterministic but chaotic dynamics} \label{subsection:problem_in_learning_stochastic_time_series_as_deterministic_dynamics_of_chaos}

Here, we explain a problem that appears in particular when stochastic time series are learned by a dynamical system.
To examine the problem, we consider the difficulty of learning of time series by RNNs for several categories of target training data sequences.

Firstly, we consider the case in which training data are given by deterministic dynamical systems including chaos, in contrast with the case of nondeterministic time series.
If the training data contain all degrees of freedom, then further states are uniquely determined by a current state.
Thus, the task of generating the time series is equivalent to the approximation of an input-output function, and it is not necessary for successful learning that the model has internal states with recurrent connections.
On the other hand, if the training data contain only a subset of the degrees of freedom, then further states cannot be determined uniquely by a current state.
However, by the Takens' embedding theorem, the training data can often be embedded in another Euclidean space, in which time evolution is deterministic, by using the method of delays.
Accordingly, even if training data realize only a subset of the degrees of freedom, we can still reduce the task to the approximation of an input-output function by reconstructing the state space.
From this fact, the training data can be learned by a model which has a contraction mapping with respect to internal states (e.g., an echo state network), because the contraction mapping plays the role of embedding past time series in the internal states.
The successful learning of Mackey-Glass chaotic time series by an echo state network is a remarkable example \cite{Jaeger2004}.

By way of contrast, nondeterministic sequences generated by a stochastic process cannot be transformed to deterministic time series by reconstructing the state space with the method of delays.
Hence, the training data of nondeterministic sequences cannot be learned in the same way as deterministic sequences.
One of the ways to output nondeterministic sequences using a dynamical system model is for the model to exhibit chaotic dynamics with respect to internal states, and for the output of the model to consist of a readout of the internal states.
If encoded information of nondeterministic sequences can be embedded in the initial states with enough precision to predict future values, nondeterminacy of the sequences can be imitated by ``sensitive dependence on initial conditions'', which was discovered by Edward Lorenz in the early 1960s.
For instance, RNNs utilizing sensitive dependence on initial conditions can learn nondeterministic sequences (e.g., an experimental study in \cite{Tani95b} or our simulation in section \ref{section:simulation}).
Learning for such models is, however, often unstable, for the following reason.
In order to train such models, a set of learnable parameters has to include initial states, because information about sequences is embedded in the initial states.
If a trained model is governed by a chaotic dynamics, almost any minute change of an initial state brings about a drastic change in the output sequence, due to the initial sensitivity of the chaotic dynamics.
Hence this initial sensitivity introduces instability into learning processes.

\subsection{Memory capacity of initial states for embedding sequence information} \label{subsection:memory_capacity_of_initial_states}

As described in section \ref{subsection:problem_in_learning_stochastic_time_series_as_deterministic_dynamics_of_chaos}, in order for a dynamical systems model utilizing sensitive dependence on initial conditions to successfully learn nondeterministic sequences, it has to be possible to embed information about sequences in the initial states with enough precision to predict future values.
However, initial states containing enough information to enable prediction is not a sufficient condition for successful learning, because there is the possibility for trained models to memorize exactly all the training data but not to acquire any chaotic dynamics.
We intuitively expect that a sufficient amount of training data will be necessary to force a more general model; i.e., one that has acquired an adequate chaotic dynamics.
We will discuss conditions for obtaining a chaotic dynamics taking into account both the amount of training data and the memory capacity of internal states.

\begin{figure}
\begin{center}
\scalebox{1.0}{\includegraphics{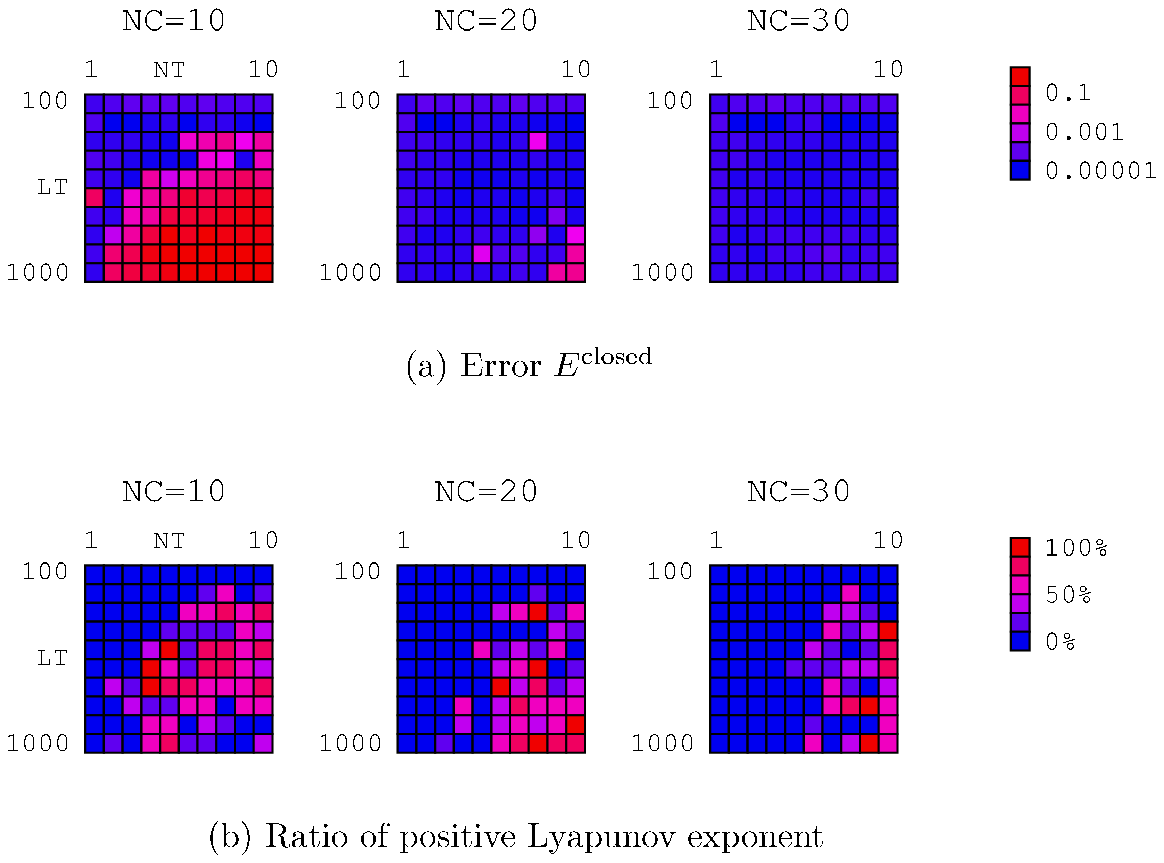}}
\caption{
Phase plots of the error $E^{\mathrm{closed}}$ and the ratio of networks whose maximum Lyapunov exponent is positive.
The upper figures display the error $E^{\mathrm{closed}}$, and the lower figures display the ratio of positive Lyapunov exponent.
In the figures, $NC$ denotes the number of context neurons in a gating network, along the vertical axis is the length of each training sequence $LT$, and along the horizontal axis is the number of training sequences $NT$.
}
\label{figure:closed_loop_error_of_the_model}
\end{center}
\end{figure}

To examine conditions for obtaining chaotic dynamics by learning for a mixture of RNN experts, we present additional simulation results.
In the following, we refer to the number of training sequences, the step length of each training sequence and the number of context neurons in a gating network as $NT$, $LT$, and $NC$, respectively.
Here, we repeat the learning simulation of the task in section \ref{subsection:learning_stochastically_switching_of_Lissajous_curves} with the transition probability $P$ set as $0.5$, with varying $NT$, $LT$ and $NC$.
In addition, in order to measure the memory capacity of initial states, we define an error $E^{\mathrm{closed}}$ for a closed-loop dynamics such that
\begin{equation}
E^{\mathrm{closed}} = \frac{1}{2Td} \sum_{n=1}^{T} ||\boldsymbol{{y}}_n^{\mathrm{teach}} - \boldsymbol{y}_{n}^{\mathrm{closed}}||^2,
\end{equation}
where $\boldsymbol{y}_n^{\mathrm{teach}}$ denotes the training data, and $\boldsymbol{y}_{n}^{\mathrm{closed}}$ denotes output of the model computed for a closed-loop dynamics with a trained initial state.
If $E^{\mathrm{closed}}$ becomes sufficiently small, the mixture of RNN experts can output sequences almost the same as the training data, namely, the training data are embedded in initial states.
Figure \ref{figure:closed_loop_error_of_the_model} shows the results of training for each parameter setting such as the number of sequences $NT$, their step length $LT$ and the number of context neurons $NC$.
In this figure, we display the average of the error $E^{\mathrm{closed}}$ and the ratio of networks whose maximum Lyapunov exponent is positive, for $10$ samples for each parameter setting.
The results in Figure \ref{figure:closed_loop_error_of_the_model} (a) clearly indicate that the error $E^{\mathrm{closed}}$ depends on both $NT$ and $LT$, and that $E^{\mathrm{closed}}$ decreases with increasing $NC$.
In particular, $E^{\mathrm{closed}}$ increases if $NT$ and $LT$ are increased in the case of $NC = 10$.
This result is due to the fact that error margins cannot be corrected because of less memory capacity in initial states.
The results in Figure \ref{figure:closed_loop_error_of_the_model} (b) indicate that chaos tends to appear if both $NT$ and $LT$ are increased.
In addition, by increasing $NC$, generating chaos tends to require a greater amount of training data.
It can be inferred that overabundant memory capacity enables the networks to memorize exactly all training data without extracting an adequate chaotic dynamics from the data.
Thus, if networks can exhibit sufficient memory capacity to embed training sequences in initial states, a small number of context neurons tends to generalize an adequate chaotic dynamics from the training data.
On the other hand, in the case of $10$ context neurons, the trained networks did not acquire a chaotic dynamics even if subjected to a great deal of training data.
This result implies that acquiring chaos tends to fail if the memory capacity of a network is insufficient to embed training sequences.
If a task demands a large amount of training data, a sufficient number of context neurons is necessary to successfully learn the task.
Therefore, these results imply that it is necessary to balance a sufficient amount of training data with the memory capacity, in order to embed the target stochastic sequences into chaos.

Finally, we explain the reason that long sequences can be encoded in initial states in our model.
It is known that the BPTT method cannot propagate effectively error signals over long-time windows, because information in the error signals decreases exponentially through the sigmoid function \cite{Bengio94}.
For instance, in studies \cite{NishimotoNN04,Tani95b} all training data learned by RNNs were of length less than $100$.
Our model, however, successfully learns training data whose length is over $10,000$.
The difference between the situation in the previous studies and our situation is brought about by the time constant $\epsilon^{g}$ in equation (\ref{equation:mixture_of_rnn_experts5}) for updating context neural states.
Since the internal potential of a context neuron decays slowly by the effect of $\epsilon^{g}$, our model can keep values of context neurons for long time.
The internal potentials of context neurons with time constant $\epsilon^{g}$ play a role similar to that of linear integrators within special units of LSTM \cite{Hochreiter97,Schmidhuber2002}.
While units of LSTM can retain values without any damping, context neurons with $\epsilon^{g}$ lose information through updating states.
Conversely, since the internal potentials of context neurons with $\epsilon^{g}$ are bounded by damping, the potential values do not diverge to infinity, and so the dynamics of our model is more stable in many situations.
As potential nonlinearity of a chaotic dynamics often forces divergence on a learning model, our model is desirable when a chaotic dynamics is learned.

\subsection{Structural stability of learning models} \label{subsection:structural_stability_of_learning_models}

In the above discussion, we demonstrated that the proposed model can have enough memory capacity to embed long sequences in initial states.
However, as we saw in the results of experiments, chaos does not always appear even if both a sufficient amount of training data and memory capacity are given and the error is made sufficiently small.
When learning models have to obtain chaotic dynamic, the problem of structural stability of the learning models appears.
The notion of structural stability was formulated by Andronov and Pontryagin \cite{Andronov1937} on the basis of the belief that a model for a natural phenomenon must preserve at least its topological character under perturbation.
Mathematically, a dynamical system is called structurally stable in the $C^1$ sense if any other dynamical system in the $C^1$ neighborhood of such a dynamical system is topologically conjugate with the dynamical system concerned.
In terms of a learning procedure using the gradient descent method, if a model is structurally stable and the learning rate is sufficiently small, the model preserves its topological character when parameters are changed in the neighborhood of a local minimum.
By contrast, in the case when a model is not structurally stable, convergence of a likelihood (or an error) does not correspond to convergence of the topological character of the model, and success or failure of learning cannot be identified until actually computing the time evolution.
Therefore, structural stability of the model is preferable from the viewpoint of learning.
However, taking into account the whole space of dynamical systems, structurally stable dynamical systems are not dense in this space.
Hence a neural network, which is a kind of universal function approximator, is not structurally stable in general.
For example, in the simulation described in Figure \ref{figure:error_kldiv_lyapunov_for_Lissajous9} the network seems not to be structurally stable, while in Figure \ref{figure:error_kldiv_lyapunov_for_reber_grammar} and Figure \ref{figure:error_kldiv_lyapunov_for_Lissajous2} the networks might be structurally stable at the end of training.
If we use a universal approximator as a learning model, it is difficult to avoid occurrence of a model that is not structurally stable.

\begin{figure}
\begin{center}
\scalebox{1.0}{\includegraphics{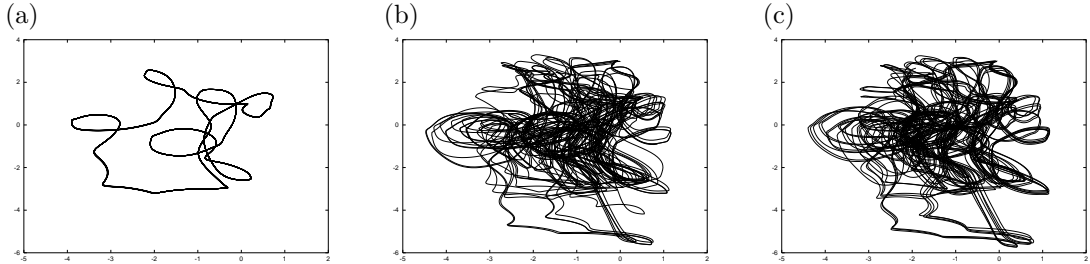}}
\caption{Trajectories of two values of $\boldsymbol{u}_{n}^{g}$ of a network with closed-loop dynamics, where the network is trained as in section \ref{subsection:memory_capacity_of_initial_states} with $10$ training sequences, the step length of each sequence being $1000$ and $30$ context neurons.
For each figure, we plotted over $10,000$ time steps after convergence of the transients.
(a) A trajectory of the network after $10,000$ learning steps (the maximum Lyapunov exponent is $-0.000054$).
(b) A trajectory of the network after $10,010$ learning steps (the maximum Lyapunov exponent is $0.000995$).
Dynamics of the network changed to chaos from a limit cycle after only $10$ learning steps.
(c) A trajectory of the network the same as in case (a), but we added Gaussian white noise with variance $\rho^2 = 0.0001$ to the input ${x}_n$ for each time step $n$.
Chaotic dynamics appears when the noise is added.
}
\label{figure:2d_context_with_noise}
\end{center}
\end{figure}

Let us consider how to take steps to cope with the appearance of a model that is not structurally stable.
One possible way is to add a negligible amount of noise to the dynamics of a trained model.
For instance, even if the dynamics of a trained model is not exactly chaotic because it is not structurally stable, the model could acquire a dynamical structure whose neighborhood contained the chaotic dynamics.
In this case, by adding low noise, the model can recure chaotic movements by making transverse intersections between stable and unstable manifolds.
In addition, if the model is structurally stable, the topological character does not change for low noise by satisfying the shadowing property \cite{Robinson99}.
As an example, Figure \ref{figure:2d_context_with_noise} displays trajectories of a network, where the network is trained in section \ref{subsection:memory_capacity_of_initial_states} and it seems not to be structurally stable.
Figure \ref{figure:2d_context_with_noise} (a) shows a trajectory of the network after $10,000$ learning steps.
In this case, the dynamics of the network is a limit cycle and the maximum Lyapunov exponent is negative.
In Figure \ref{figure:2d_context_with_noise} (b), however, the dynamics of the network changes to chaos from a limit cycle after only $10$ learning steps.
Next, we added Gaussian white noise with variance $\rho^2 = 0.0001$ to the dynamics of the network having a limit cycle, and plotted a trajectory in Figure \ref{figure:2d_context_with_noise} (c).
Consequently, a dynamics similar to that of case (b) can be reconstructed by adding the noise.
It is shown that a neighborhood of the network contains a chaotic system, and adding noise restores the network dynamics to the chaotic system while respecting the original topology.

\begin{figure}
\begin{center}
\scalebox{1.0}{\includegraphics{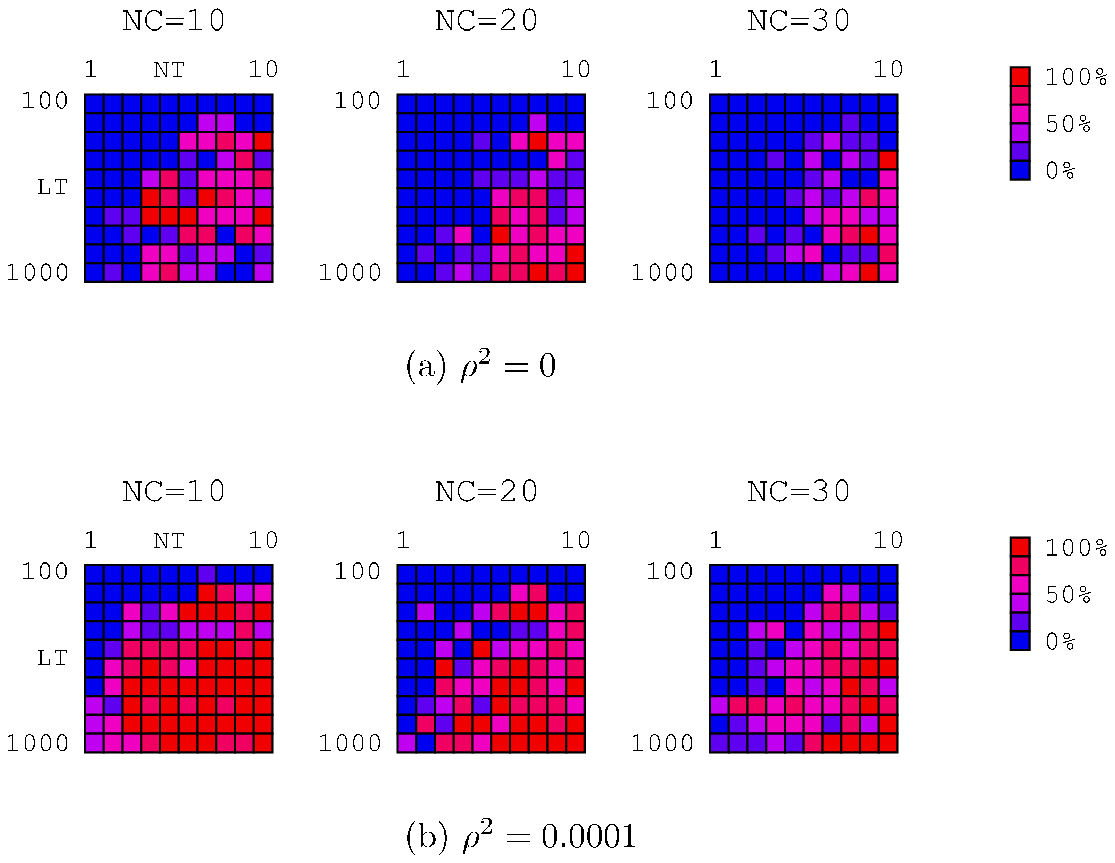}}
\caption{
The ratio of networks generating nonperiodic output sequences.
Here, Gaussian white noise with variance $\rho^2$ is added to the input ${x}_n$ for each time step.
(a) Phase plots corresponding to the case of $\rho^2 = 0$, namely, without noise.
(b) Phase plots corresponding to the case of $\rho^2 = 0.001$.
In these figures, $NC$ denotes the number of context neurons in the gating network, the vertical axis is the length of each training sequence $LT$, and the horizontal axis is the number of training sequences $NT$.
}
\label{figure:periodicity_of_the_model}
\end{center}
\end{figure}

In Figure \ref{figure:periodicity_of_the_model}, to evaluate the effect of low noise we display ratios of networks generating nonperiodic output sequences, where the networks were trained in section \ref{subsection:memory_capacity_of_initial_states}, and Gaussian white noise with variance $\rho^2$ was added to the input ${x}_n$ for each time step $n$.
Figures \ref{figure:periodicity_of_the_model} (a) and (b) display the results for variances $\rho^2 = 0$ and $\rho^2 = 0.001$, respectively.
Nonperiodicity of output sequences implies that the dynamics is chaotic or quasi-periodic.
To evaluate periodicity of output sequences, we labeled a left and right circle in output trajectories (see Figure \ref{figure:2d_teach_and_output_for_Lissajous2}) as a symbol $\mathrm{L}$ and $\mathrm{R}$, respectively.
Forming symbolic sequences like $\mathrm{LRLLRRLR} \cdots$ from output trajectories, we computed to determine periodicity of the symbolic sequences over a period of $100$.
Comparing Figure \ref{figure:2d_teach_and_output_for_Lissajous2} (a) with (b), it can be seen that the proportion of networks generating nonperiodic sequences is increased by adding the noise.
On the other hand, in the region corresponding to a small amount of training data, output is periodic even if noise is added.
It turned out that by adding noise, output of a network becomes nonperiodic if the network is trained by using enough training data, whereas output is periodic if we use less training data.
In addition, adding noise does not result only in nonperiodic sequences being generated, because there is a region in which output is periodic even if noise is added.
Thus, this result implies that in the neighborhood of a trained network subjected to a sufficient amount of data, there exists an appropriate chaotic dynamics.
Hence, we can classify trained networks into three classes from the results about their dynamics: (1) those that have a structurally stable and periodic dynamics, (2) those that have a structurally stable and chaotic dynamics and (3) those that are structurally unstable but with a nearby system having a chaotic dynamics.
Models with low noise can generate chaotic time series and imitate nondeterministic characteristics underlying the training data not only in case (2) but also in case (3).

\section{Conclusion} \label{section:conclusion}

This paper has shown that a mixture of RNN experts model can acquire the ability to generate sequences combining multiple primitives.
By virtue of training of the model, each expert learns a primitive sequence pattern, and a gating network learns a switching rule between winners to make decisions about which expert to use as a winner on each occasion.
Our simulation experiments showed that the model imitates stochastic switching between the multiple primitives in terms of a chaotic dynamics utilizing sensitive dependence on initial conditions.
It was also demonstrated that a self-organized chaotic system can reconstruct the probability of primitive switching as observed in the training data.
We have also considered two essential problems concerning memory capacity and structural stability of the model.
Through analysis of the network dynamics, we have inferred that a sufficient amount of training data, balanced with the memory capacity of the network is a necessary condition for embedding stochastic time series into a chaotic dynamical system.
We have also inferred that even if the dynamics of the model is not exactly chaotic because of structural instability, adding a negligible amount of noise into the model can enable stably reconstruction of stochastic time series.
The model can be applied to prediction and generation of time-series, especially when the time series to be learned are potentially able to be represented as a composition of several primitives, and there is no required pre-knowledge of the primitives or a composition rule for them.

\bibliography{bibliography07}

\end{document}